\documentclass{article}

\usepackage{spconf}
\usepackage{amsmath,amssymb,amsfonts}
\usepackage{bm}
\usepackage{bbm}
\usepackage{graphicx}
\usepackage{booktabs}
\usepackage{tabularx}
\usepackage{mathrsfs}


\def\Xb{{\bm X}}
\def\xb{{\bm x}}
\def\yb{{\bm y}}
\def\rb{{\bm r}}
\def\db{{\bm d}}
\def\Db{{\bm D}}
\def\Eb{{\bm E}}

\newcommand{\Exp}{\mathbb{E}}
\newcommand{\Ind}{\mathbbm{1}}
\newcommand{\FDP}{\mathrm{FDP}}

\newcommand{\FDR}{\mathrm{FDR}}
\newcommand{\one}{\mathbf{1}}
\newcommand{\sgn}{\operatorname{sign}}
\newcommand{\inner}[2]{\left\langle #1,#2\right\rangle}

\newcolumntype{Y}{>{\raggedleft\arraybackslash}X}
\newcolumntype{Z}{>{\raggedright\arraybackslash}X}

\title{SCALABLE VARIABLE SELECTION UNDER PREDICTOR DEPENDENCE WITH ADAPTIVE VIRTUAL DUMMIES}

\name{Taulant Koka and Michael Muma%
\thanks{The work of T. Koka and M. Muma has been funded by the ERC Starting Grant ScReeningData, Grant No. 101042407.}}

\address{Robust Data Science Group\\
Technische Universität Darmstadt}

\begin{document}

\ninept
\maketitle

\begin{abstract}
Reliable high-dimensional variable selection requires scalable error-controlling methods. The Terminating-Random Experiments (T-Rex) selector estimates the false discovery rate (FDR) by aggregating early-terminated forward-selection paths in which predictors compete with synthetic dummies. We address two remaining challenges: i) predictor dependence can bias dummy–predictor competition; ii) computation is wasted on recomputing terms shared across experiments. Building on memory-efficient virtual dummies, which sequentially sample projections from their exact conditional law, we estimate the conditional Gaussian law of inactive predictors and map draws onto the remaining sphere radius. A shared lazy Gram cache computes the response product and each requested Gram column once across experiments. Simulations across three covariance structures show that uniform spherical dummies can exceed the target FDR, whereas the proposed method empirically controls FDR. Caching yields more than a sixfold speedup, and the method remains feasible with $100\:000$ predictors, where competing FDR-controlling methods become computationally impractical.
\end{abstract}


\section{Introduction}
\label{sec:intro}

High-dimensional variable selection seeks to determine which of many candidate predictors are associated with a response. In genome-wide association studies, for example, few genetic variants among a large candidate set may influence a trait. Reliable selection therefore requires controlling the amount of irrelevant reported variables, commonly formalized through the false discovery rate (FDR) \cite{Benjamini1995}. Let $\widehat{\mathcal A}\subseteq\{1,\ldots,p\}$ denote the selected set and $\mathscr N$ the null-variable indices. The false discovery proportion and FDR are
\begin{equation}
 \FDP
 =
 \frac{|\widehat{\mathcal A}\cap\mathscr N|}
 {\max\{|\widehat{\mathcal A}|,1\}},
 \qquad
 \FDR
 =
 \Exp\!\left[\FDP\right].
 \label{eq:fdr}
\end{equation}

When $p\geq n$, full-model $p$-values required by Benjamini--Hochberg and Benjamini--Yekutieli are unavailable without structural assumptions or pre-screening, which can reduce power \cite{Benjamini1995,Benjamini2001}. Knockoffs reproduce dependence in synthetic controls \cite{Barber2015,Cands2018}, but Model-X requires a joint predictor model; Gaussian implementations estimate and factorize a covariance matrix.

The Terminating-Random Experiments (T-Rex) framework repeatedly competes real predictors with synthetic dummy variables in forward selectors and analyzes the ensemble of terminated random experiments to compute a conservative FDR estimator \cite{Machkour2025}.  Each of $B$ random experiments augments the design with $L$ dummies and traces the path of a forward selector, such as orthogonal matching pursuit (OMP) \cite{pati1993orthogonal} or least angle regression (LARS) \cite{Efron2004}, until $T$ dummies enter. Variables that precede the dummies sufficiently often are retained, and their relative selection frequencies enter the dummy-based FDR calibration. However, the guarantee for independent dummies does not transfer unchanged to general dependence. The dependency-aware extension (T-Rex+DA) addresses this limitation by grouping correlated predictors and penalizing selection frequencies across correlation thresholds \cite{Machkour2025_Dependent}. In addition to computing $B$ terminated forward paths, each involving $L$ dummy competitors, T-Rex+DA requires constructing the groups and searching the additional threshold grid, which further increases the runtime. Furthemore, the penalization of the selection frequencies can make the selector conservative. At very large scales, explicitly storing and processing an $n\times L$ dummy matrix can itself dominate resources.

Virtual dummies \cite{Koka2026Virtual} alleviate the dummy-related cost by generating only the coordinates required along a forward-selection path and sampling the full dummy vector only if it is selected. They reproduce the joint and conditional laws of explicit rotationally invariant dummies without materializing the entire dummy matrix. However, independent spherical dummies still ignore anisotropic null competition, and the $B$ paths still repeat products with the unchanged real design. These remaining statistical and computational limitations motivate two contributions:
\par\noindent\emph{(i)}
Sphere-constrained adaptive Gaussian dummies, referred to as Sphere AG, fit a conditional Gaussian model to inactive-real projections and map each draw onto the remaining sphere radius. This is an empirical dependence adaptation rather than a new FDR theorem.
\par\noindent\emph{(ii)}
Shared lazy Gram caching computes the response product once and each requested real Gram column only once across all paths. It avoids the full Gram matrix and preserves the paths for fixed and random inputs. Code and experiment scripts are available from the authors' github page.\footnote{\texttt{https://github.com/taulantkoka}}

\section{Dependence-adaptive virtual dummies}
\label{sec:method}

For i.i.d.\ centered Gaussian designs, standardized true-null columns and independent dummies sampled from the uniform probability distribution on the centered unit sphere share the rotationally invariant law needed for fair T-Rex competition; their projection distributions are characterized in \cite{Khokhlov2006}. Under dependence, inactive-real projections become anisotropic and path dependent, whereas spherical dummy projections do not.

\begin{figure}[!t]
\centering
\includegraphics[width=0.76\columnwidth]
{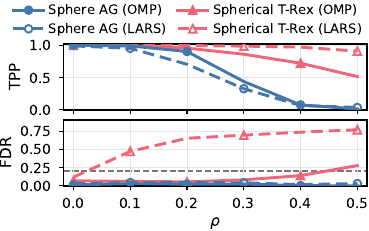}
\caption{Sphere AG and ordinary spherical T-Rex under grouped-factor dependence over $100$ paired trials. TPP is shown above FDR; OMP uses solid curves and LARS dashed curves. Bars show one standard error, and the dark dashed horizontal line marks the target FDR $0.2$.}
\label{fig:sphere-uniform}
\end{figure}

Fig.~\ref{fig:sphere-uniform} illustrates the resulting failure mode under grouped-factor dependence at SNR one. For both OMP and LARS, spherical T-Rex increasingly exceeds the target FDR as $\rho$ grows. Sphere AG remains empirically below the target, with a power loss under strong dependence. Section~\ref{sec:experiments} gives the complete simulation design.

This comparison identifies the dummy projection law as the component to adapt. We retain the published T-Rex stopping, voting, and FDP-estimation rules so that the change concerns only dependence-aware competition. This isolation supports a direct empirical comparison, but does not extend the standard T-Rex proof.

We next describe the virtual representation underlying Sphere AG, then the real-design work shared by all random experiments.

Let $\Xb=(\xb_1,\ldots,\xb_p)\in\mathbb R^{n\times p}$ have centered unit-norm columns, and let $\yb\in\mathbb R^n$ be a nonzero centered response. The predictors, response, and dummies $\Db=(\db_1,\ldots,\db_L)$ lie in
\[
 H=\{\bm u\in\mathbb R^n:\one^\top\bm u=0\},
 \qquad
 m=n-1.
\]

Virtual-dummy selection maintains an adaptive basis beginning with $\bm e_1=\yb/\|\yb\|_2$. Immediately before step $k$, let $\Eb_k=(\bm e_1,\ldots,\bm e_k)$. If $\bm v_k$ is the real or dummy column selected at that step, the next basis direction is
\begin{equation}
 \bm e_{k+1}
 =
 \frac{(\bm I-\Eb_k\Eb_k^\top)\bm v_k}
 {\|(\bm I-\Eb_k\Eb_k^\top)\bm v_k\|_2}.
 \label{eq:basis-update}
\end{equation}
Rather than storing $\db_\ell$, the virtual procedure generates only its coordinates $\alpha_{\ell,k}=\inner{\db_\ell}{\bm e_k}$ as new basis directions become available. Thus, after $k$ directions it stores $O(Lk)$ scalar coordinates rather than $O(nL)$ dummy entries. If $\db_\ell$ is selected, its unrevealed component in $H\cap\operatorname{span}(\Eb_k)^\perp$ is completed from its conditional law. Under rotational invariance, this reproduces the explicit-dummy selection law exactly \cite{Koka2026Virtual}.

This representation removes the explicit dummy matrix, but not the cost of the $B$ forward paths. Every experiment still queries the same $\Xb^\top\yb$ and, when real variable $j$ enters, the same Gram column $\Xb^\top\xb_j$. Section~\ref{sec:gram} removes this repeated real-design work through shared lazy Gram caching after we address the dummy law.

In experiment $b$, let $\mathcal C_b(T)$ contain the real variables entering before the $T$th dummy. The relative occurrence of a real variable $j$ is
\[
 \Phi_{T,L}(j)
 =
 B^{-1}
 \sum_{b=1}^{B}
 \Ind\{j\in\mathcal C_b(T)\},
\]
where $\Ind\{\cdot\}$ is the indicator function. For voting level $v$, the T-Rex selector reports
\[
 \widehat{\mathcal A}_{T,L}(v)
 =
 \{j:\Phi_{T,L}(j)>v\}
\]
and searches its admissible $(T,v)$ pairs for the largest selected set whose estimated FDP does not exceed the target level \cite{Machkour2025}. We retain this calibration and modify only the conditional law used to generate $\alpha_{\ell,k}$.

\subsection{Sphere-constrained adaptive Gaussian dummies}

When direction $\bm e_k$ is revealed, let $\mathcal J_{k-1}$ be the inactive real indices and let $n_{k-1}=|\mathcal J_{k-1}|$. For $j\in\mathcal J_{k-1}$, define $\bm z_{j,k}=\Eb_k^\top\xb_j$. Sphere AG estimates
\begin{equation}
\begin{split}
 \widehat{\bm\mu}_k
 &=
 \frac{1}{n_{k-1}}
 \sum_{j\in\mathcal J_{k-1}}\bm z_{j,k},
 \\
 \widehat{\bm\Sigma}_k
 &=
 \frac{1}{n_{k-1}-1}
 \sum_{j\in\mathcal J_{k-1}}
 (\bm z_{j,k}-\widehat{\bm\mu}_k)
 (\bm z_{j,k}-\widehat{\bm\mu}_k)^\top.
\end{split}
\label{eq:moments}
\end{equation}

For dummy $\db_\ell$, let
\[
 \bm a_{\ell,k-1}
 =
 (\alpha_{\ell,1},\ldots,\alpha_{\ell,k-1})^\top,
 \qquad
 R_{\ell,k-1}^2
 =
 1-\|\bm a_{\ell,k-1}\|_2^2.
\]
Partitioning into the previous $k-1$ coordinates and the current coordinate yields:
\[
 \widehat{\bm\mu}_k
 =
 \begin{bmatrix}
  \widehat{\bm m}_k\\
  \widehat\nu_k
 \end{bmatrix},
 \qquad
 \widehat{\bm\Sigma}_k
 =
 \begin{bmatrix}
  \widehat{\bm A}_k & \widehat{\bm c}_k\\
  \widehat{\bm c}_k^\top & \widehat q_k
 \end{bmatrix}.
\]
Conditioning the $k$th fitted projection coordinate on the previously generated coordinates $\bm a_{\ell,k-1}$ gives
\begin{equation}
\begin{split}
 \widetilde\mu_{\ell,k}
 &=
 \widehat\nu_k
 +
 \widehat{\bm c}_k^\top\widehat{\bm A}_k^{-1}
 (\bm a_{\ell,k-1}-\widehat{\bm m}_k),
 \\
 \widehat\sigma_k^2
 &=
 \widehat q_k
 -
 \widehat{\bm c}_k^\top
 \widehat{\bm A}_k^{-1}\widehat{\bm c}_k.
\end{split}
\label{eq:conditional-gaussian}
\end{equation}

To preserve unit norm, the Gaussian draw is combined with an estimate of the complementary energy. For $k<m$, set
\begin{equation}
 \tau_k^2
 =
 \frac{
 \sum_{j\in\mathcal J_{k-1}}
 [1-\|\bm z_{j,k}\|_2^2]_+
 }
 {n_{k-1}(m-k)},
 \label{eq:complement-scale}
\end{equation}
and draw
\(
 G_{\ell,k}
 \sim
 \mathcal N(\widetilde\mu_{\ell,k},\widehat\sigma_k^2)\:
 \) and \(\:
 W_{\ell,k}
 \sim
 \tau_k^2\chi^2_{m-k}.
\)
Then, update
\begin{equation}
 U_{\ell,k}
 =
 \frac{G_{\ell,k}^2}{G_{\ell,k}^2+W_{\ell,k}},
 \quad
 \alpha_{\ell,k}
 =
 R_{\ell,k-1}\sgn(G_{\ell,k})\sqrt{U_{\ell,k}},
 \label{eq:adaptive-update}
\end{equation}
with
\(
 R_{\ell,k}^2
 =
 R_{\ell,k-1}^2(1-U_{\ell,k}).
\)

Here $G_{\ell,k}$ is a latent directional score and $W_{\ell,k}$ represents energy in the unrevealed directions. Their ratio maps the fitted Gaussian score into the remaining radius.

\noindent\textbf{Proposition 1.}
The update in \eqref{eq:adaptive-update} satisfies
\(
 \alpha_{\ell,k}^2+R_{\ell,k}^2=R_{\ell,k-1}^2.
\)

\noindent If $\widetilde\mu_{\ell,k}=0$ and $\widehat\sigma_k^2=\tau_k^2$, then \eqref{eq:adaptive-update} coincides with the spherical virtual-dummy construction, and
\[
 U_{\ell,k}
 \sim
 \operatorname{Beta}
 \left(
 \frac{1}{2},
 \frac{m-k}{2}
 \right).
\]

\noindent\textit{Proof.}
The radius identity follows directly from \eqref{eq:adaptive-update}. Under the stated conditions, $G_{\ell,k}^2/\tau_k^2\sim\chi_1^2$ and $W_{\ell,k}/\tau_k^2\sim\chi_{m-k}^2$ independently, yielding the beta law. \hfill$\square$

If $\db_\ell$ is selected, its remaining component is sampled from the uniform probability distribution on the corresponding sphere in $H\cap\operatorname{span}(\Eb_k)^\perp$. Together with Proposition 1, this ensures that every realized dummy is centered and has unit norm. Outside the isotropic special case, Sphere AG is an empirical approximation to the path-dependent projection distribution of the inactive real predictors.

\noindent\textbf{Scope of the statistical claim.}
Proposition 1 guarantees dummy geometry, not FDR control. Since the fitted moments depend on the design and selected path, Sphere AG lacks the null--dummy exchangeability used by the standard T-Rex proof. Its FDR behavior is therefore evaluated empirically.

\section{Shared lazy Gram caching}
\label{sec:gram}

Across the $B$ paths, every experiment accesses the same real design $\Xb$. The vector $\Xb^\top\yb$ is therefore common to all paths, and every path that selects real predictor $j$ requires the same Gram column
\(
 \bm g_j=\Xb^\top\xb_j.
\)
Rather than evaluating $\Xb^\top\Xb$ in full, shared lazy Gram caching computes $\bm g_j$ only upon its first request by any experiment and reuses it thereafter. This reuse is especially effective in combination with virtual dummies, where experiment $b$ does not require the augmented design $(\Xb,\Db^{(b)})$. With explicitly generated dummies, the dummy-related Gram columns would depend on $\Db^{(b)}$ and could not be reused across experiments. Under virtual dummies, the real design remains the cost-dominating shared object.

Let $\mathcal A_b^{\mathrm r}$ and $\mathcal A_b^{\mathrm d}$ be the active real and dummy indices in experiment $b$. For OMP, the real-design residual correlations satisfy
\begin{equation}
 \Xb^\top\rb^{(b)}
 =
 \Xb^\top\yb
 -
 \sum_{j\in\mathcal A_b^{\mathrm r}}
 \widehat\beta_j^{(b)}\bm g_j
 -
 \sum_{\ell\in\mathcal A_b^{\mathrm d}}
 \widehat\theta_\ell^{(b)}
 \Xb^\top\db_\ell^{(b)}.
 \label{eq:cached}
\end{equation}
Here $\widehat\beta_j^{(b)}$ and $\widehat\theta_\ell^{(b)}$ are the current active-path coefficients. The response product and real Gram columns in the first two terms persist across experiments. The final term is experiment specific: its virtual-dummy projections are generated only along the realized path. LARS likewise uses the cached real Gram columns \cite{Efron2004}.

We compute $\Xb^\top\yb$ once and maintain a thread-safe lazy Gram cache $j\mapsto\bm g_j$. The first request for index $j$ computes the corresponding column, while later requests reuse it. A memory budget limits the number of retained columns and prevents formation of the full $p\times p$ Gram matrix. Because cached and direct multiplication return the same inner products, caching is a computational change only: it leaves the inferential output unchanged in exact arithmetic and whenever rounding does not alter the candidate ordering.

Let $\kappa_b$ denote the path length in experiment $b$, and let $M_G$ count first-time real-column requests across all experiments. Every experiment selects $T$ dummy directions before termination. The dominant real-design cost changes from
\(
 \mathcal O\left(
 np\sum_{b=1}^{B}\kappa_b
 \right)
\)
to
\begin{equation}
 \mathcal O\left(
 np\left[1+M_G+BT\right]
 +
 p\sum_{b=1}^{B}\kappa_b^2
 \right).
 \label{eq:complexity}
\end{equation}
The initial $np$ term computes $\Xb^\top\yb$, $npM_G$ computes distinct requested real Gram columns, and $npBT$ covers experiment-specific dummy directions. The remaining term updates the active systems along the paths. With sufficient cache capacity, $M_G$ is the number of distinct real predictors selected by at least one experiment, so reuse grows when paths repeatedly request the same predictors.

\begin{table}[!t]
\caption{Runtime in seconds, mean $\pm$ standard error over $100$ paired repetitions on a 64 GB, 10-core Apple M1 Max. Cache scaling uses ten threads; the end-to-end comparison uses one.}
\label{tab:runtime}
\centering

\setlength{\tabcolsep}{1.4pt}
\renewcommand{\arraystretch}{0.92}

{\textbf{Lazy Gram caching}}\\[1pt]

\begin{tabularx}{\columnwidth}{@{}ZYYY@{}}
\multicolumn{4}{@{}l}{\textit{(a) Joint dimension scaling
($B=20$)}}\\
\toprule
$(n,p)/10^3$ & Direct & Lazy & Speedup\\
\midrule
$(2.5,2.5)$
& $1.077\pm0.031$
& $0.249\pm0.005$
& $4.32\pm0.08$\\
$(5,5)$
& $3.485\pm0.018$
& $0.777\pm0.005$
& $4.49\pm0.03$\\
$(7.5,7.5)$
& $7.803\pm0.078$
& $1.682\pm0.016$
& $4.65\pm0.03$\\
$(10,10)$
& $13.216\pm0.062$
& $2.837\pm0.012$
& $4.66\pm0.02$\\
\bottomrule

\addlinespace[3pt]
\multicolumn{4}{@{}l}{\textit{(b) Scaling in the number of experiments
($n=p=5000$)}}\\
\toprule
$B$ & Direct & Lazy & Speedup\\
\midrule
$5$
& $1.148\pm0.003$
& $0.483\pm0.003$
& $2.38\pm0.01$\\
$10$
& $1.733\pm0.012$
& $0.544\pm0.004$
& $3.20\pm0.02$\\
$20$
& $3.329\pm0.010$
& $0.755\pm0.003$
& $4.41\pm0.02$\\
$40$
& $6.843\pm0.031$
& $1.256\pm0.005$
& $5.45\pm0.03$\\
$80$
& $13.752\pm0.107$
& $2.230\pm0.018$
& $6.18\pm0.03$\\
\bottomrule
\end{tabularx}

\vspace{4pt}

{\textbf{End-to-end selector comparison}}\\[-1pt]
{$n=1000$, $p=2500$, $B=20$}\\[1pt]

\begin{tabular*}{\columnwidth}
{@{\extracolsep{\fill}}lrr@{}}
\toprule
Method & OMP & LARS\\
\midrule
Sphere AG
& $0.450\pm0.008$
& $0.721\pm0.019$\\
Spherical T-Rex
& $0.393\pm0.007$
& $0.733\pm0.027$\\
T-Rex+DA
& $2.181\pm0.032$
& $2.689\pm0.098$\\
Gaussian MX / MX+
& \multicolumn{2}{c}{$68.382\pm0.186$ (shared)}\\
\bottomrule
\end{tabular*}
\end{table}

\begin{figure*}[!t]
\centering
\includegraphics[width=0.88\textwidth]
{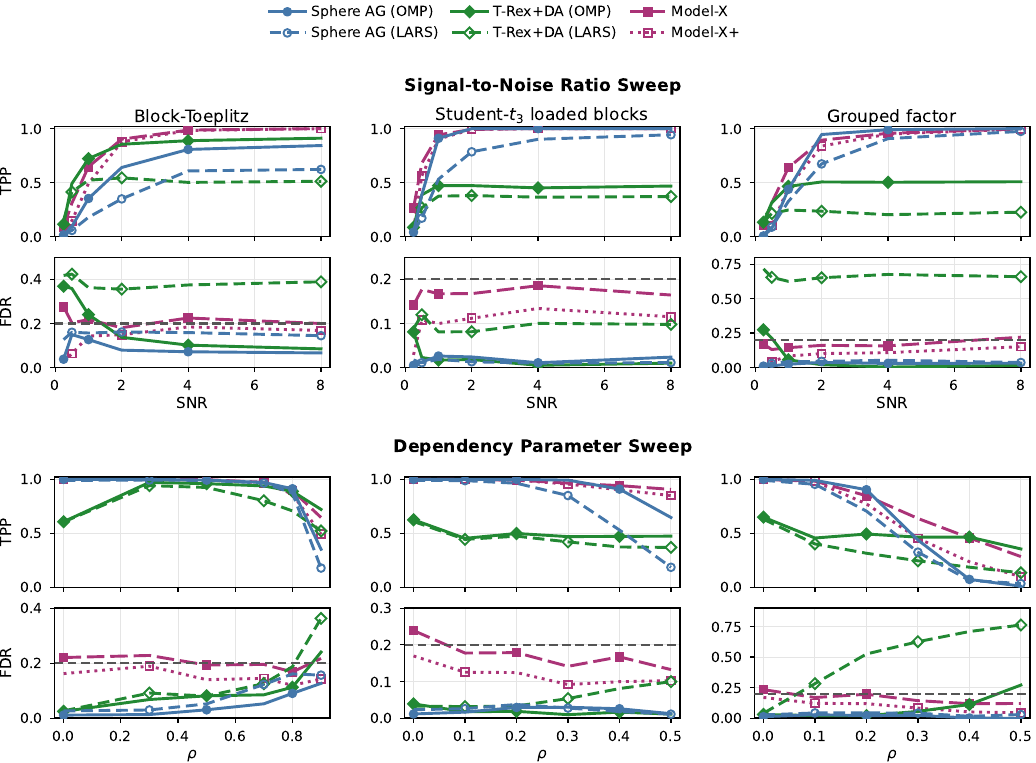}
\caption{TPP and FDR across SNR and dependence over $100$ paired trials. Sphere AG and T-Rex+DA use OMP/LARS; Model-X variants use Lasso.}
\label{fig:benchmark}
\end{figure*}

\section{Experiments}
\label{sec:experiments}

Unless otherwise stated, we use $n=500$, $p=1000$, $p_1=10$, $B=20$, and $L=3p$. Columns are centered and normalized to unit norm. We draw $\mathcal A$ uniformly without replacement from all supports of size $p_1$, set $\beta_j=p_1^{-1/2}$ for $j\in\mathcal A$ and zero otherwise, and generate
\[
 \yb=\Xb\bm\beta+\bm\epsilon,
 \qquad
 \bm\epsilon\sim\mathcal N(\bm0,\sigma^2\bm I),
 \qquad
 \sigma^2=\frac{\operatorname{Var}(\Xb\bm\beta)}{\mathrm{SNR}}.
\]
At each operating point, all methods use the same $100$ data sets, supports, responses, and random seeds. After the spherical baseline study in Fig.~\ref{fig:sphere-uniform}, the main benchmark compares Sphere AG and T-Rex+DA \cite{Machkour2025_Dependent}, each with OMP and LARS, against Gaussian Model-X and Model-X+ knockoffs using a Ledoit--Wolf covariance estimate \cite{LedoitWolf2004} and a Lasso statistic. The target FDR is $\alpha=0.2$.

Model-X and Model-X+ share one draw and statistic and differ only in the threshold offset. Since the predictor covariance is estimated with $p>n$, these are empirical plug-in comparators, not oracle Model-X procedures supplied with the known covariance.

Table~\ref{tab:runtime} contains two runtime comparisons. The upper block measures shared lazy Gram caching alone using Sphere AG with OMP, $L=3p$, $T=5$, and block-Toeplitz dependence with $\rho=0.9$. Panel (a) varies $n=p$ with $B=20$, while panel (b) varies $B$ with $n=p=5000$. Across all $900$ paired runs, selected sets and occurrence matrices agree exactly. The speedup reaches $4.66\times$ at $n=p=10000$ and grows from $2.38\times$ at $B=5$ to $6.18\times$ at $B=80$. The lower block compares complete selectors at $(n,p)=(1000,2500)$, SNR one, $\rho=0.9$, and one thread. Sphere AG is close in runtime to spherical T-Rex and substantially faster than T-Rex+DA. Gaussian Model-X and Model-X+ share one knockoff draw and Lasso fit, but their shared computation remains considerably slower.

SNR sweeps fix block-Toeplitz dependence at $\rho=0.9$, Student-$t_3$-driven block loadings at $\rho=0.4$, and grouped-factor dependence at $\rho=0.3$. We also sweep $\rho$ while fixing the SNR at one. Both block models use blocks of size $100$. Block-Toeplitz predictors have within-block covariance $\rho^{|j-j'|}$ and are independent across blocks. For the Student-$t$ loaded-block model, a standardized $z\sim t_3$ gives
\vspace{-5pt}
\[
 \lambda_z^2
 =
 \frac{\rho}{2}
 +
 \frac{\rho}{2}
 \frac{z^2}{1+z^2},
 \qquad
 \sgn(\lambda_z)
 =
 \sgn(z).
\vspace{-5pt}
\]
Within each block, predictor $j$ is generated as $X_j=\lambda_j f+\sqrt{1-\lambda_j^2}\,\epsilon_j$. The grouped-factor model uses ten groups and decomposes each predictor as $X_j=\sqrt{0.25\rho}\,f_0+\sqrt{0.75\rho}\,f_{g(j)}+\sqrt{1-\rho}\,\epsilon_j$, with independent standard Gaussian factors and noise.

Fig.~\ref{fig:benchmark} shows Sphere AG below the target for both solvers, with TPP rising with SNR and falling under grouped-factor dependence. T-Rex+DA's FDR rises with dependence, especially for LARS in that model; Model-X+ stays below the target, whereas Model-X exceeds it at several points. Sphere AG therefore trades power for empirical FDR stability.

\begin{table}[!t]
\caption{Large-scale grouped-factor study with $n=1000$, $p=100\:000$, $p_1=10$, $\rho=0.3$, $B=20$, $L=3p$, and calibrated $T$ over $100$ trials per SNR. SNR cells give FDR/TPP; maximum standard errors are $.008/.040$. Time is averaged over all $300$ runs per method. Model-X and T-Rex+DA could not be run for this setting.}
\label{tab:large-scale}
\centering
\scriptsize
\begin{tabular*}{\columnwidth}{@{\extracolsep{\fill}}lcccc@{}}
\toprule
& \multicolumn{3}{c}{SNR} & Time\\
\cmidrule(lr){2-4}
Method & $1$ & $2$ & $4$ & (s)\\
\midrule
Sphere AG, OMP
& $.007/.341$ & $.028/.812$ & $.041/.930$ & $4.94$\\
Sphere AG, LARS
& $.012/.125$ & $.011/.364$ & $.016/.609$ & $3.74$\\
Spherical, OMP
& $.038/.966$ & $.014/1.00$ & $.014/1.00$ & $4.05$\\
Spherical, LARS
& $.885/.998$ & $.885/1.00$ & $.884/1.00$ & $27.0$\\
\bottomrule
\end{tabular*}
\end{table}

Table~\ref{tab:large-scale} shows Sphere AG below the target with increasing power. Spherical OMP also stays below it, but spherical LARS has FDR near $.885$ and selects about $89$ variables. The tested Model-X implementation would need $80$ GB RAM and could not be run on the $64$ GB machine, while the hierarchical clustering step for grouping in the T-Rex+DA is not viable. 

\section{Conclusion}

Predictor dependence exposes a practical hurdle: spherical dummies can misrepresent null competition, while dependency-aware alternatives can be costly. Sphere AG adapts dummies by learning path-conditional projections from inactive predictors while preserving the spherical geometry resulting from standardization. It yields stable empirical FDR, with some power loss in challenging highly dependent regimes.

Virtual sampling and the shared lazy Gram cache preserve paths, scale the adaptive method, and deliver more than a sixfold speedup. Sphere AG remains close to spherical T-Rex in runtime, outpaces tested T-Rex+DA and Model-X, and runs at $p=100\:000$ in seconds, where those two are not viable. 

While the proposed pooled Gaussian fit may capture broadly shared dependence, it may miss a small subset of strongly response-aligned nulls. Future work will therefore focus on localized or mixture-based projection laws with finite-sample FDR guarantees.
\clearpage
\bibliographystyle{IEEEbib}
\bibliography{references}

\end{document}